\RequirePackage{fix-cm}
\documentclass[smallextended]{svjour3}       
\smartqed  

\usepackage{graphicx} 
\usepackage{lineno,bm,float,soul,amsfonts,amsmath,siunitx}
\usepackage{xcolor,hyperref}
\usepackage{empheq}
\usepackage{multirow}
\usepackage{rotating}
\usepackage{adjustbox}

\usepackage{adjustbox}
\usepackage{stmaryrd}

\newcommand{\T}{^{\mbox{\tiny T}}}

\newcommand{\B}[1]{{\bm #1}}

\begin{document}

\title{Poincar\'e Maps with the Theory of Functional Connections
}


\author{A. K. de Almeida Jr   \href{https://orcid.org/0000-0002-9488-4462}{\includegraphics[scale=0.5]{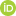}}     \and
        D. Mortari \href{https://orcid.org/0000-0003-0787-4547}{\includegraphics[scale=0.5]{ORCID-iD_icon_16x16.png}} 
}


\institute{A. K. de Almeida Jr  \at
              CFisUC, Departamento de Física, Universidade de Coimbra, 3004-516 Coimbra, Portugal \\
              Atlar Innovation, Aveiro, Portugal\\
              \email{allan.junior@inpe.br}           
           \and
           D. Mortari \at
              Aerospace Engineering, Texas A\&M University, College Station TX, 77843.\\
              \email{mortari@tamu.edu} 
}

\date{Received: date / Accepted: date}

\maketitle

\begin{abstract}
Poincaré maps play a fundamental role in nonlinear dynamics and chaos theory, offering a means to reduce the dimensionality of continuous dynamical systems by tracking the intersections of trajectories with lower-dimensional section surfaces. Traditional approaches typically rely on numerical integration and interpolation to detect these crossings, which can lead to inaccuracies and computational inefficiencies, especially in systems characterized by long-term evolution or sensitivity to initial conditions. This work presents a novel methodology for constructing Poincaré maps based on the Theory of Functional Connections (TFC). The constrained functionals produced by TFC yield continuous and differentiable representations of system trajectories that exactly satisfy prescribed constraints. The computation of Poincaré maps is formulated as either an initial value problem (IVP) or a boundary value problem (BVP). For IVPs, initial conditions are embedded into the functional, and the intersection time with a specified section surface is determined via root-finding techniques. We demonstrate linear convergence to the Taylor series, thereby enabling accurate interpolation without resorting to numerical integration or external optimization for short time intervals. For BVPs, periodicity conditions are encoded to identify periodic orbits such as families of Lyapunov and Distant Retrograde Orbits in a Circular Restricted Three-Body Problem context. Furthermore, by enforcing partial periodic constraints, we show how to construct first recurrence maps with selective control over specific components of position and/or velocity. The methodology is also extended to non-autonomous systems, demonstrated through applications to the Bicircular Biplanar Four-Body Problem. The proposed approach achieves machine-level accuracy with modest computational effort, eliminating the need for variable transformations or iterative integration schemes with adaptive step-sizing. The results illustrate that TFC offers a powerful and efficient alternative framework for constructing Poincaré maps, computing periodic orbits, and analyzing complex dynamical systems, particularly in astrodynamical contexts.
\keywords{Theory of Functional Connections \and Poincaré Maps \and astrodynamics \and periodic orbits}
\end{abstract}

\section{Introduction}

Poincar\'e sections are fundamental tools in the study of dynamical systems and chaos theory, providing an effective means to analyze the qualitative behavior of complex, high-dimensional systems. Instead of examining the entire continuous trajectory in phase space, one studies a discrete set of points corresponding to the intersections of the trajectory with a lower-dimensional hypersurface, known as the surface of section. This dimensionality reduction enables the visualization and identification of dynamic behaviors such as regular motion, characterized by structured, repetitive point sets, and chaotic motion, indicated by irregular or fractal-like distributions. Moreover, Poincar\'e maps facilitate the identification of periodic orbits and invariant structures such as limit cycles, strange attractors, and manifolds. In particular, periodic orbits are of great interest as they often reveal hidden symmetries and underlying structure in perturbed dynamical systems. In astrodynamics, periodic orbits can be strategically exploited to reduce the fuel required for station-keeping and transfer maneuvers, thereby enhancing mission efficiency.

The \textit{Theory of Functional Connections} (TFC) is a recently introduced mathematical framework \cite{U-ToC} that performs linear functional interpolation. It constructs analytical functionals, called \textit{constrained functionals}, that analytically satisfy a prescribed set of linear constraints, thereby transforming constrained problems into unconstrained ones. This transformation significantly simplifies the solution of differential equations by restricting the whole solution space to the subspace of functions that inherently satisfy the constraints. TFC has been successfully applied in various scientific and engineering domains, including the solution of differential equations \cite{Leake2019,Leake2020,florion2,Wang2024}, optimal control and trajectory optimization \cite{Johnston2020,Li2021,Wang2022}, particle physics \cite{Floriononame1,Schiassi2022}, biological modeling \cite{Xu2020,Daryakenari2024}, and geodesy \cite{Mortari2022}, among many others. Its applicability can be further extended via coordinate transformations \cite{tfcvariables} and domain mapping \cite{bijective}.

In astrodynamics, TFC has demonstrated notable effectiveness in problems involving satellite characterization, orbit determination \cite{akajtfcnelder}, Earth–Moon transfers \cite{Earth2Moon,akajtangential,Campana2024,tfc_connecting_earth_moon_2025}, and the computation of periodic orbits \cite{DEALMEIDA2023102068}. Initially applied to solve initial value problems (IVPs) for linear, differential equations using least-squares techniques \cite {LDE}, the method was later extended to nonlinear problems \cite{NDE}. The capability to obtain a final estimate with machine error accuracy level was adopted by TFC to solve differential equations with oscillating and/or chaotic solutions.

This work investigates an alternative formulation of the IVP using TFC with two constraints, as outlined in \cite{NDE}. Using a specified set of support functions, we demonstrate linear convergence of the constrained functional with respect to a Taylor expansion of the solution around the initial point. We demonstrate that this convergence is guaranteed independently of the free function, thereby eliminating the need for external procedures to generate an initial guess (such as the Runge-Kutta integrator adopted in \cite{NDE}) to start a subsequent optimization procedure. This demonstration improves the robustness and efficiency of the method for short time intervals. Furthermore, we take advantage of the novel constrained functional to solve for its root and evaluate crossings with a surface of section, independently of external interpolation procedures, as explained next.

Since the constrained functional yields a continuous, analytic solution over each time interval, it serves as a highly accurate interpolation method, inherently satisfying the governing differential equation at numerous points within the interval. This property is particularly advantageous for computing Poincar\'e maps, where one must determine the precise time at which a trajectory intersects a given surface of section. Traditional methods typically integrate in time until the sign of a section function changes, and then apply an interpolation between the final two points to estimate the intersection. This approach suffers from potential numerical inaccuracies and high computational costs. Although integrating in spatial variables is an alternative \cite{Henon1982}, it often requires complex variable transformations.

This paper proposes a TFC-based solution to the Poincar\'e map problem formulated as either an IVP or a boundary value problem (BVP). The technique capitalizes on the continuous, differentiable structure of the constrained functional to evaluate accurate intersections with section surfaces, either analytically (when using low-order free functions) or numerically (via standard root-finding methods). We present first a formulation based on IVPs in Sec.~\ref{sec:IVP}. After that, we present BVP formulations in Sec.~\ref{sec:bvp} for computing periodic orbits, tailoring the application shown in \cite{periodic} to define a surface of section and extending it to partially periodic orbits. The BVP-based techniques presented in this paper enable us to determine the boundaries such that a single recurrence is mapped on the surface of section for a specified time. We also take advantage of such a specification to decrease the dimensionality of time-dependent systems by analyzing it in a time-discrete domain, extending then the technique presented in this paper to \textit{stroboscopic maps} in Sec. \ref{sec:strob}. These methods improve the accuracy and computational efficiency of constructing Poincar\'e maps and provide a flexible framework for embedding a wide range of physical constraints relevant to astrodynamical applications.

\section{Using the constrained functional to intersect the surface of section}

Consider a differential equation of the second order subject to constraints. We use a three-dimensional column vector formulation for the position and its derivatives, where the independent variable is the time $t$, the dependent variable is $\B{r}$, its first derivative with respect to time is $\dot{\B{r}}$, and the second derivative with respect to time is $\ddot{\B{r}}$. 
The second-order differential equations can then be written,
\begin{equation}\label{eq:eqmotion}
    \ddot{\B{r}} = \B{a} (\B{r}, \dot{\B{r}}),
\end{equation}
where $\B{a}$ is an acceleration function depending on position and velocity, $\B{r}$ and  $\dot{\B{r}}$. The problem is then to find the solution $\B{r} (t)$ of Eq.~(\ref{eq:eqmotion}) for the time $t$ in the interval $\{0,T\}$, considering the specified constraints.

Using three-dimensional vectors, the constrained functional can be obtained from \cite{U-TFC,tfc_solarsail} ($\eta$ formulation)
\begin{equation}\label{eq:generating2}
    \B{r} = \B{g} + E \, \B{s},
\end{equation}
where $\B{g} = \B{g} (t)$ is a column vector of free functions of order $3\times 1$, $E$ is a constant matrix of order $3\times k$, and $\B{s} = \B{s} (t)$ is a vector of support functions of order $k \times 1$, where $k$ is the number of constraints of the problem.

A linear system can be generated by applying Eq.~(\ref{eq:generating2}) by enforcing the constraints. This system is then analytically solved for $E$. The constrained functional, thus obtained using this procedure, has the property of embedding the constraints into a single expression.

On the other hand, a surface of section can be defined by the equation
\begin{equation}\label{eq:ss}
    S(\B{r},\dot{\B{r}}) = 0,
\end{equation}
which in general has one less dimension with respect to the dimensions of $\B{r}$ and $\dot{\B{r}}$ combined.
In this paper, we take advantage of the constrained functional derived through TFC to search for crossings of the trajectory in the phase space with this surface. We investigate several different formulations to solve this problem using several forms of the constrained functional, analyzing their advantages and limitations.

\section{Surface crossings via IVP}
\label{sec:IVP}

The initial value problem is characterized here by two constants: the position and velocity evaluated at the starting time. This problem is described here through the following two constraints:
\begin{equation}\label{eq:constr1}
\begin{cases}
    \B{r} (t_0) = \B{r}_0,  \\ 
    \dot{\B{r}} (t_0) = \B{v}_0.
\end{cases}
\end{equation}
where $\B{r} (0)$ and $\dot{\B{r}} (0)$ are the position $\B{r} (t)$ and velocity $\dot{\B{r}} (t)$ vectors applied at $t=t_0$, while the initial conditions are $\B{r}_0$ and $\B{v}_0$.

Assuming the support function as $\B{s}=\{s_1(t),s_2(t)\}\T$, the expression shown in Eq.~(\ref{eq:generating2}) is applied to the constraints shown in Eq.~(\ref{eq:constr1}). Solving for $E$ and applying the solution into Eq.~(\ref{eq:generating2}), the constrained functional is derived as
\begin{align}\label{eq:cf1}
\B{r} (t) =& \B{g} (t)+\frac{s_2(t) \left(s_1(t_0) \left(\B{v}_0-\dot{\B{g}}_0\right)+(\B{g}_0-\B{r}_0)\dot{s}_1(t_0)\right)}{s_1(t_0)\dot{s}_2(t_0)-s_2(t_0)\dot{s}_1(t_0)}\\
&\quad +\frac{s_1(t) \left(s_2(t_0) \left(\dot{\B{g}}_0-\B{v}_0\right)+(\B{r}_0-\B{g}_0)\dot{s}_2(t_0)\right)}{s_1(t_0)\dot{s}_2(t_0)-s_2(t_0)\dot{s}_1(t_0)} \nonumber
\end{align}
where $\B{g}_0=\B{g} (t_0)$ and $\dot{\B{g}}_0=\dot{\B{g}} (t)\big|_{t_0}$.

\subsection{Convergence through the choice of support functions}

A fundamental requirement when applying TFC for optimization is that the functions $\B{s}$ and $\B{g}$ must be linearly independent, as discussed in \cite{NDE}. Beyond this necessary condition, this section investigates the choice of support functions that ensures the technique remains independent of external procedures.

We demonstrate that if an appropriate form of $\B{s}$ is selected, then, unlike conventional series expansions around $t_0$, the convergence of the constrained functional depends linearly on time. This property enables the selection of a sufficiently small time interval, eliminating the need for an improved initial guess $\B{g} (t)$ in optimization routines such as nonlinear least squares.

To analyze this behavior, we consider a first-order approximation of the free function $\B{g} (t)$ expanded around $t = t_0$:
\begin{equation}\label{eq:g_exp2}
    \B{g} (t) = \B{g}_0 + \dot{\B{g}}_0 (t - t_0) + \mathcal{O}\big((t - t_0)^2\big).
\end{equation}
Substituting Eq.~\eqref{eq:g_exp2} into the expression for the constrained functional (Eq.~\eqref{eq:cf1}) yields
\begin{align}\label{eq:cf1a}
   \B{r} (t) &= \B{g}_0 + \dot{\B{g}}_0 (t - t_0) \\
   &\quad + \frac{s_2(t) \left[s_1(t_0) (\B{v}_0 - \dot{\B{g}}_0) + (\B{g}_0 - \B{r}_0) \dot{s}_1(t_0)\right]}{s_1(t_0) \dot{s}_2(t_0) - s_2(t_0) \dot{s}_1(t_0)} \nonumber \\
   &\quad + \frac{s_1(t) \left[s_2(t_0) (\dot{\B{g}}_0 - \B{v}_0) + (\B{r}_0 - \B{g}_0) \dot{s}_2(t_0)\right]}{s_1(t_0) \dot{s}_2(t_0) - s_2(t_0) \dot{s}_1(t_0)} + \mathcal{O}\big((t - t_0)^2\big). \nonumber
\end{align}

Now, we define the support functions as $s_1(t) = k_1$ and $s_2(t) = k_2 t$, where $k_1, k_2 \in \mathbb{R}$. With this choice, Eq.~\eqref{eq:cf1a} simplifies to
\begin{equation}\label{eq:cf1b}
    \B{r} (t) = \B{r}_0 + \B{v}_0 (t - t_0) + \mathcal{O}\big((t - t_0)^2\big),
\end{equation}
which matches the first-order Taylor series expansion of $\B{r} (t)$ about $t = t_0$.

This result indicates that if the support functions are defined as the first-order natural polynomials, i.e., $\B{s} = \{k_1, k_2 t\}\T$, the constrained functional exhibits linear convergence with respect to a Taylor expansion around $t = t_0$. Crucially, this linear convergence is independent of the free function $\B{g} (t)$. The same result would be obtained by selecting $\B{g} (t) = \B{0}$. This selection transforms functional interpolation into interpolation as obtained by the selected support functions. provided by Taylor in Eq. \eqref{eq:cf1b}. Thus, convergence is ensured for values of $t$ near $t_0$, regardless of the initial guess for the coefficients defining $\B{g} (t)$ (see the Appendix for details on these coefficients). Consequently, there is no need to employ external procedures, such as using a Runge-Kutta integrator to update the unknowns (see matrix $L$ in appendix) for initializing nonlinear least squares optimization, as done in \cite{NDE}, provided a small time interval is adopted.

For the above reasons, in this work, the support functions $\B{s} = \{1, t\}\T$, are selected. This ensures, within a sufficiently small interval, linear convergence relative to a Taylor expansion around $t = t_0$, independent of the initial guess for $\B{g} (t)$. With this definition, the constrained functional in Eq.~\eqref{eq:cf1} becomes
\begin{equation}\label{eq:cf2}
    \B{r} \big(t, \B{g} (t)\big) = \B{g} (t) + (\B{r}_0 - \B{g}_0) + \left(\B{v}_0 - \dot{\B{g}}_0\right)(t - t_0),
\end{equation}
which satisfies the constraint conditions in Eq.~\eqref{eq:constr1} for any choice of the free function $\B{g} (t)$.

\subsection{Iterative procedure and numerical parameters}\label{sec:parameters}

We aim to solve Eq.~\eqref{eq:eqmotion} over the time interval $t \in [0, T]$. When the final time $T$ is sufficiently small, the solution can be obtained directly over the entire interval using the procedure outlined in the previous sections. However, if $T$ is relatively large, direct numerical convergence may not be guaranteed due to the local nature of the convergence properties. In such cases, the time interval $[0, T]$ is partitioned into smaller subintervals of length $T_s$, over which the solution is computed iteratively.

In this iterative framework, each time step spans the interval $t \in [t_0, T_s]$, where $t_0$ denotes the initial time of the current subinterval. The step size $T_s$ is selected such that higher-order (quadratic and beyond) terms in the expansion with respect to $(t - t_0)$ can be neglected. According to the method presented earlier, for each subinterval of length $T_s$, the solution $\B{r} (t)$ and its derivative $\dot{\B{r}} (t)$ are computed. These values at $t = T_s$, denoted by $\B{r}_i(T_s)$ and $\dot{\B{r}}_i(T_s)$ for the $i$-th iteration, serve as the initial conditions for the subsequent $(i+1)$-th step. That is,
\begin{equation}
    \B{r}_i (T_s) = \B{r}_{i+1} (t_0), \qquad \text{and} \qquad \dot{\B{r}}_i (T_s) = \dot{\B{r}}_{i+1} (t_0),
\end{equation}
where these values define the constraints as specified in Eq.~\eqref{eq:constr1}, representing the initial value problem (IVP) for each iteration.

Although a sufficiently small value of $T_s$ ensures convergence, it may significantly increase the computational cost. Since the technique is coupled with a nonlinear optimization routine (e.g., a nonlinear least squares procedure), it is often possible to employ a larger step size $T_s$ while still maintaining convergence. This is because the free function $\B{g} (t)$ can be numerically adjusted to compensate for discrepancies between the approximated solution given by Eq.~\eqref{eq:cf1b} and the actual solution of the system, which depends on the true specific force $\B{a} (t)$ for $t > t_0$.

Thus, $T_s$ should be selected to balance computational efficiency and convergence accuracy: large enough to reduce the total number of steps, yet sufficiently small to ensure the validity of the linear approximation and the effectiveness of the optimization.

Moreover, the numerical implementation introduces two additional parameters:
\begin{itemize}
\item $N$: the number of discretization points in time $t$ used to evaluate the constrained functional and optimize the free function;

\item $m$: the number of basis functions employed to construct the free function $\B{g} (t)$.
\end{itemize}

The parameters $T_s$, $N$, and $m$ must be carefully chosen to achieve an appropriate trade-off between convergence, accuracy, and computational efficiency. This choice may depend on the characteristics of the specific force function $\B{a} (t)$. Further details regarding the numerical procedure for optimizing $\B{g} (t)$ via nonlinear least squares are provided in the Appendix.

\subsection{Solving for the surface of section}\label{sec:solving_surface}

In this section, we present a method for utilizing the constrained functional defined in Eq. \eqref{eq:ss}. Since both the position vector, $\B{r} (t)$, and its time derivative, $\dot{\B{r}} (t)$, vary with time, the primary challenge lies in identifying the precise time, denoted as $t_c$, at which the surface of section condition is satisfied, namely when Eq.~\eqref{eq:ss} holds. Note that the constrained functional in Eq. \eqref{eq:cf2} is applicable for both forward ($t > t_0$) and backward ($t < t_0$) time propagation.

The free function, $\B{g} (t)$, may be constructed using elementary mathematical expressions, such as a truncated series of orthogonal polynomials (see Appendix). This approach yields an analytical formulation for $\B{r} (t)$ which, under lower-order truncations, may be invertible, thereby enabling an explicit expression for time as a function of position. In favorable cases, the root of the equation $S \big(\B{r} (t), \dot{\B{r}} (t)\big) = 0$ can be determined analytically and integrated directly into the computational algorithm.

However, due to the localized nature of the approximation and the fact that the interpolation error remains of order $\mathcal{O}\big((t - t_0)^2\big)$, independent of the specific choice of $\B{g} (t)$, an accurate initial guess for $t_0$ near the intersection is required. These errors may be further mitigated by optimizing $\B{g} (t)$ through the numerical procedure outlined previously. When higher-order truncations of $\B{g} (t)$ are employed, analytical inversion may become impractical. In such scenarios, root-finding algorithms, including Newton's method or the bisection method, may be utilized to solve $S \big(\B{r} (t), \dot{\B{r}} (t)\big) = 0$, where $\B{r} (t)$ and $\dot{\B{r}} (t)$ adhere to the formulation in Eq.~\eqref{eq:cf2}. The computed values of position and velocity at the intersection time, $t_c$, are subsequently incorporated into the construction of the Poincaré map.

\subsection{Comparisons with other techniques}

A direct method for determining the intersection of a trajectory in phase space with the surface of section involves integrating the trajectory and subsequently adjusting the final integration step to revert to a dynamical state before the crossing, utilizing a reduced time step. The approach outlined in this section, formulated as an initial value problem (IVP), eliminates the need for successive integrations or adaptive step-size reductions near the surface. Instead, it leverages the constrained functional to resolve the intersection either analytically or numerically. This methodology is capable of achieving machine-level accuracy while maintaining low computational cost, provided that the numerical parameters specified in Sec.~\ref{sec:parameters} are appropriately calibrated. When employing a numerical solution, determining the root of a continuous function of a single variable proves to be significantly more efficient than direct numerical integration of the equations of motion with iterative time updates.

An auxiliary interpolation procedure could be applied between two points, one positioned on either side of the surface of section. Notably, the constrained functional, as a continuous formulation, inherently provides a highly accurate representation in the vicinity of $t = t_0$. Consequently, this technique obviates the necessity for external interpolation procedures, thereby reducing computational overhead while achieving machine-level accuracy with ease.

This method constitutes a practical alternative to the transformation set proposed in \cite{Henon1982} for analytically addressing the intersection problem from an initial state proximate to the crossing. Specifically, the problem may be solved using either the analytical approach, where an invertible expression for $t(\B{r})$ is available, or the numerical approach, which employs Newton's method, as detailed in Sec.~\ref{sec:solving_surface}.

Additionally, subsequent sections introduce complementary techniques based on the constrained functional that enable the exact determination of surface crossings through the imposition of constraints. These techniques are formulated as boundary value problems (BVPs).

\section{Surface crossings via BVP}\label{sec:bvp}

This section presents a methodology for identifying the first recurrence at a prescribed time. While Section~\ref{sec:IVP} introduced a technique to determine the time of crossing using an initial value problem (IVP) formulation, the boundary value problem (BVP) approach developed here takes the specified time as input and instead focuses on identifying the corresponding boundary conditions. The term first recurrence refers to a pair of points on the surface of section such that one is the image of the other under the system's flow, with the additional constraint that they are separated by an exact, predefined time interval. This approach not only reduces the problem's dimensionality by restricting the analysis to the surface of section but also enables a simultaneous discretization of time into specified intervals.

\subsection{Periodic solutions}\label{sec:bvp1}

Periodic trajectories in phase space can be characterized by the following constraints:
\begin{equation}\label{eq:constr}
    \begin{cases}
        \B{r}(0) = \B{r}(T),  \\ 
        \dot{\B{r}}(0) = \dot{\B{r}}(T),
    \end{cases}
\end{equation}
where $\B{r}(t)$ and $\dot{\B{r}}(t)$ denote the position and velocity vectors, respectively, and $T$ is the trajectory period. Note that the initial conditions $\B{r}(0)$ and $\dot{\B{r}}(0)$ are not prescribed.

Assuming the support functions are selected as $\B{s} = \{t, t^2\}\T$, the combination of Eq.~\eqref{eq:generating2} with the constraints in Eq.~\eqref{eq:constr} yields a linear system. Solving this system for $E$ leads to the following expression for the constrained functional:
\begin{equation}\label{eq:ce2c}
    \B{r}(t) = \B{g}(t) + \dfrac{t}{2 T} \bigg[(t - T) (\dot{\B{g}}_0 - \dot{\B{g}}_f) +2 (\B{g}_0 - \B{g}_f)\bigg],
\end{equation}
where $\B{g}_0 = \B{g}(0)$, $\B{g}_f = \B{g}(T)$, $\dot{\B{g}}_0 = \dot{\B{g}}(t)\big|_{t=0}$, and $\dot{\B{g}}_f = \dot{\B{g}}(t)\big|_{t=T}$. Since Eq.~\eqref{eq:ce2c} satisfies the constraints in Eq.~\eqref{eq:constr} for any choice of the free function $\B{g}(t)$, it produces solutions that are periodic with period $T/\lambda$, where $\lambda \in \mathbb{N}^+$ is the period of the periodic solution represented in the Poincaré map built with the surface of section defined in Sec.~\ref{sec:ssdef}.

It is important to highlight that the initial (or final) position and velocity are not explicitly imposed; instead, the values $\B{r}(0)$ and $\dot{\B{r}}(0)$ (or equivalently $\B{r}(T)$ and $\dot{\B{r}}(T)$) are obtained through the numerical convergence of the free function $\B{g}$. As a consequence, the formulation in Eq.~\eqref{eq:constr} may admit multiple solutions, or in some cases, no solution. Nevertheless, the solution space under these constraints is broader compared to the case where initial (or final) conditions are fully specified. This increased flexibility often improves the convergence behavior of optimization procedures employed to identify periodic trajectories in phase space.

\subsubsection{Defining a surface of section}
\label{sec:ssdef}

In addition to facilitating the search for periodic orbits, the constrained functional defined in Eq.~\eqref{eq:ce2c} yields an exact solution at the initial (or final) time, i.e., for $t = 0$ or $t = T$. These values can be employed to define a surface of section. Specifically, this surface passes through the point $\B{r}(0) = \B{r}_0$ and is orthogonal to the velocity vector $\dot{\B{r}}(0) = \dot{\B{r}}_0$. Hence, the surface of section shown in Eq.~\eqref{eq:ss} becomes, 
\begin{equation}\label{eq:surface_def}
    (\B{p}_s - \B{r}_0) \T \dot{\B{r}}_0 = 0,
\end{equation}
where $\B{p}_s$ denotes a generic point lying on the surface.

Given the surface of section defined in Eq.~\eqref{eq:surface_def}, the constraints introduced in Eq.~\eqref{eq:constr} naturally induces a first recurrence map, since they map the point $(\B{r}(0), \dot{\B{r}}(0))$ on the surface onto itself as $(\B{r}(T), \dot{\B{r}}(T))$. This mapping is valid exclusively for periodic orbits, as only such trajectories satisfy the recurrence condition implied by the constraints. To generalize this framework for the study of non-periodic trajectories, alternative constraints based on a prescribed surface of section will be introduced in Sec.~\ref{sec:exact_surface}.

\subsubsection{Application: periodic orbits in the CRTBP}

In the Circular Restricted Three-Body Problem (CRTBP), the Earth and Moon revolve in circular orbits around their common barycenter with constant angular velocity $\omega$. By adopting a rotating reference frame centered at the barycenter, the equation of motion for a spacecraft is given by \cite{symon}
\begin{equation}
	\label{eq:crtbp}
	\frac{\text{d}^2\B{r}}{\text{d}t^2} = -2\B{\omega} \times \frac{\text{d}\B{r}}{\text{d}t} - \B{\omega} \times \left(\B{\omega} \times \B{r} \right) - \frac{\mu_e}{r_e^3} \B{r}_e - \frac{\mu_m}{r_m^3} \B{r}_m,
\end{equation}
where $\B{r} = \{x, y, z\}\T$ is the position vector of the spacecraft relative to the barycenter, $\B{r}_e$ and $\B{r}_m$ denote the position vectors of the spacecraft relative to the Earth and Moon, respectively, with norms $r_e$ and $r_m$. The constants $\mu_e$ and $\mu_m$ are the gravitational parameters of the Earth and Moon, and the angular velocity vector of the rotating frame is $\B{\omega} = \omega \B{k}$, with $\B{k}$ being a unit vector along the $z$-axis.
The values of the parameters adopted in this research are the same as those used by \cite{simo1995book,YAGASAKI2004313,Yagasaki2004,topputo2013optimal} shown in Table \ref{tab:parameters}, where $R$ is the distance between Earth and Moon.
\begin{table}[ht]
	\centering
	\begin{tabular}{lll}
		\hline
		$R$ & $3.84405000 \times 10^{8}$ & m \\[0.5ex]
		$R_s$ & $1.49460947424915\times 10^{11}$ & m \\[0.5ex]
		$\mu_e$ & $3.975837768911438\times10^{14}$ & m$^3$/s$^2$  \\[0.5ex]
		$\mu_m$ & $4.890329364450684\times10^{12}$ & m$^3$/s$^2$ \\[0.5ex]
		$\mu_s$ & $1.3237395128595653 \times 10^{20}$ & m$^3$/s$^2$ \\[0.5ex] 
		$\omega$ & $2.66186135\times 10^{-6}$ & s$^{-1}$ \\[0.5ex]
		$\omega_s$&$-2.462743433827215\times 10^{-6}$ & s$^{-1}$ \\[0.5ex]
		\hline
	\end{tabular}
    \caption{Values of the parameters for the Sun-Earth-Moon system \cite{simo1995book}.}
	\label{tab:parameters}
\end{table}

As an application example, we employ the constrained functional defined in Eq.~\eqref{eq:ce2c} to identify periodic orbits in the CRTBP. In this context, the acceleration $\B{a}$ corresponds to the right-hand side of Eq.~\eqref{eq:crtbp}. For simplicity, we restrict our attention to the planar case, where $z = 0$.

Since the CRTBP admits multiple solutions, the initial guess strongly influences the specific solution to which the optimization procedure converges. To generate an appropriate initial guess for the free function $\B{g}(t)$, we employ an external nonlinear least squares method. This guess is constructed such that the constrained functional yields a trajectory approximating a circular path centered near the $L_1$ libration point. For sufficiently small radii, the solution converges to the $L_1$ equilibrium point itself, which trivially satisfies the periodicity constraints in Eq.~\eqref{eq:constr}. As the radius increases, the solution transitions to planar Lyapunov orbits, Distant Retrograde Orbits (DROs) around the Moon, and eventually to Earth-centered periodic orbits, in that order.

\begin{figure}
    \centering\includegraphics[scale=0.38]{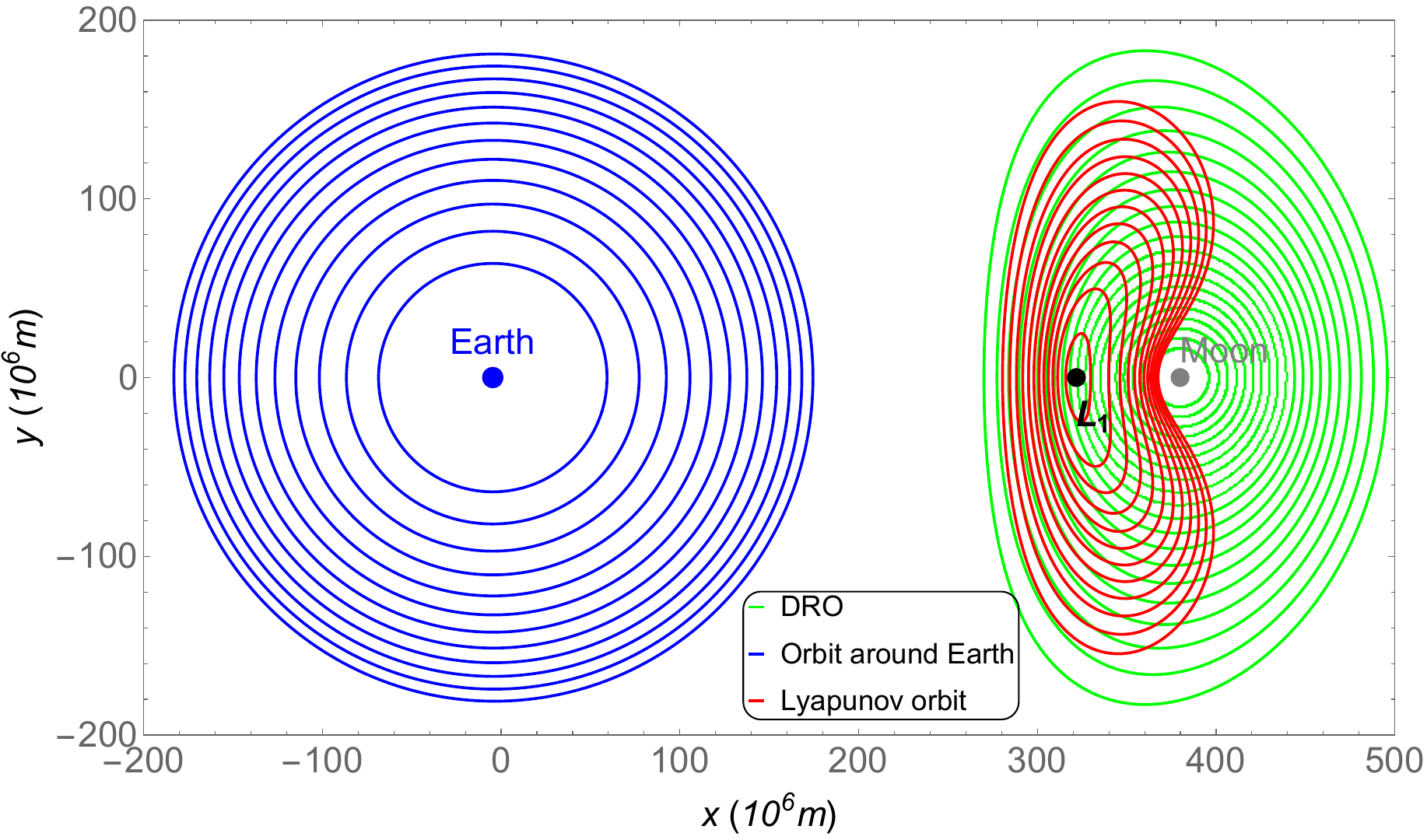}
    \caption{Planar periodic solutions in the CRTBP: Lyapunov orbits are shown in red for periods ranging from 12 days (innermost) to 23 days (outermost); Distant Retrograde Orbits (DROs) around the Moon are shown in green for periods between 2 and 21 days; and Earth-centered periodic orbits are depicted in blue for periods from 2 to 13 days. The period interval between successive orbits is 1 day. These solutions were obtained via numerical convergence using the constrained functional in Eq.~\eqref{eq:ce2c}.}
    \label{fig:lyap1}
\end{figure}

Using this approach, convergence to a periodic solution, satisfying the constraints of Eq.~\eqref{eq:constr}, is consistently achieved. Because the initial and final states are not fixed, the solution space remains large, thereby increasing the likelihood of successful convergence (see Appendix for implementation details). Starting from the described initial guess for $\B{g}(t)$, the numerical procedure converged to a family of periodic solutions. By applying a continuation method, incrementally varying the period $T$ using the previously obtained free function as a starting point, we generated the family of solutions shown in Fig.~\ref{fig:lyap1}.

In Fig.~\ref{fig:lyap1}, the red family represents planar Lyapunov orbits with periods ranging from 12 days (closest to $L_1$) to 23 days (farthest). The green family corresponds to DROs around the Moon for periods between 2 and 21 days, while the blue family depicts Earth-centered periodic orbits with periods from 2 to 13 days. The interval between successive orbits in each family is 1 day. These periodic solutions of the CRTBP are obtained using the constrained functional defined in Eq.~\eqref{eq:ce2c}.

\subsection{Partial periodic solution}\label{sec:exact_surface}

To construct a first recurrence map, also known as a Poincar\'e map, we examine intersections of the trajectory with the surface given by the $x$-$z$ plane. In this case, Eq.~(\ref{eq:ss}) becomes
\begin{equation}\label{eq:surfacey}
    y = 0
\end{equation}
considering only crossings in a single velocity direction. These intersections can be enforced precisely through the following constraints,
\begin{equation}\label{eq:constr2}
\begin{cases}
    y(0) = 0, \\
    y(T) = 0.
\end{cases}
\end{equation}

By applying rectangular coordinates in Eq.~\eqref{eq:generating2}, the \( y \)-component of the constrained functional satisfying the conditions in Eq.~\eqref{eq:constr2} is given by
\begin{equation}\label{eq:cf3}
    y(t) = g_y(t) - g_y(0) + \big(g_y(0) - g_y(T)\big)\dfrac{t}{T},
\end{equation}
where \( g_y \) denotes the \( y \)-component of the free function \( \B{g} \). Since the \( x \)- and \( z \)-components are unconstrained, their constrained functionals remain unaltered, i.e., \( x(t) = g_x(t) \) and \( z(t) = g_z(t) \).

Together, these components define the constrained functional representing a first recurrence map, which maps the initial point \( \B{r}(0) \) on the section \( y = 0 \) to the point \( \B{r}(T) \) on the same section after a fixed time \( T \).

\subsubsection{Planar case study and symmetric solutions}
\label{sec:planarsymmetryc}

\begin{figure}
	\centering
    \includegraphics[scale=0.30]{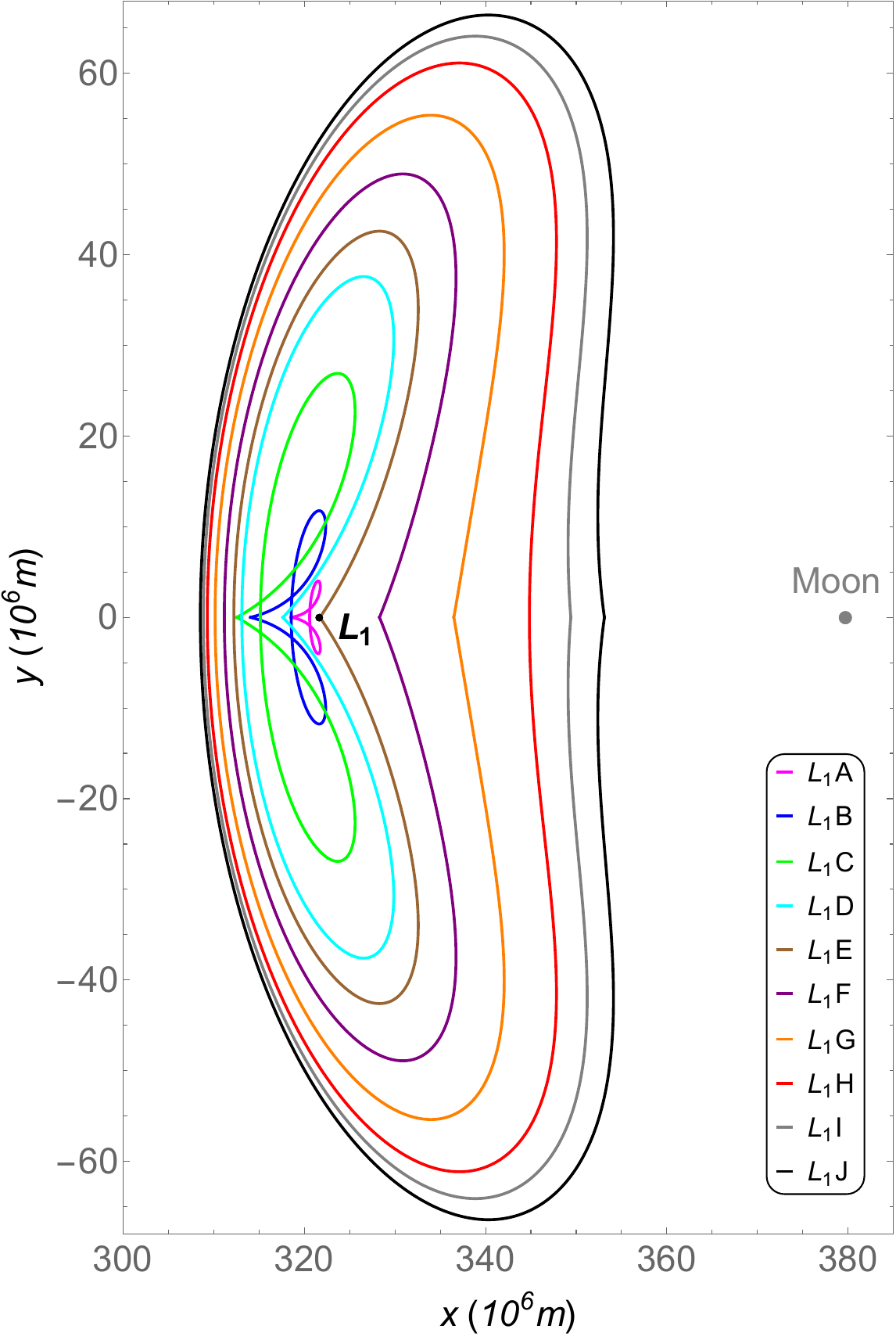}
    \includegraphics[scale=0.33]{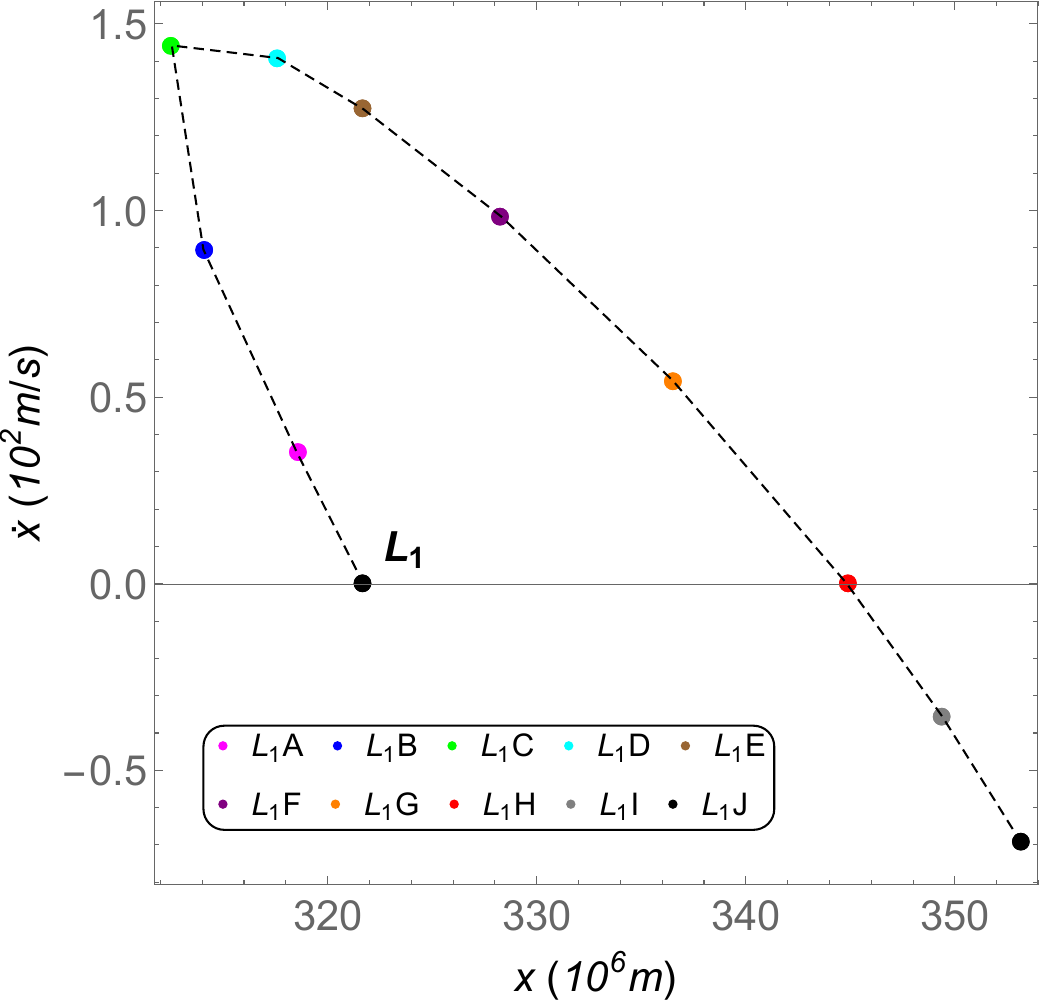}
	\caption{Orbits for $T=13.75$ days obtained using the constrained functional shown in Eq.~(\ref{eq:cf3}) is shown in the left side. The corresponding pairs $x$ and $\dot{x}$ evaluated at $t=0$ obtained from the technique are shown on the right side. Besides satisfying the constraints shown in Eq.~(\ref{eq:constr2}), they also satisfy the conditions $x(0)=x(T)$, $\dot{x} (0) =-\dot{x} (T)$, and $\dot{y} (0) = \dot{y} (T)$. The solution $L_1$H (in red) corresponds to the Lyapunov orbit, representing a particular case where $\dot{x} (0) =-\dot{x} (T)=0$. The lines are drawn to guide the eyes.}
		\label{fig:igrl1}
\end{figure}

\begin{figure}
		\centering
            \includegraphics[scale=0.33]{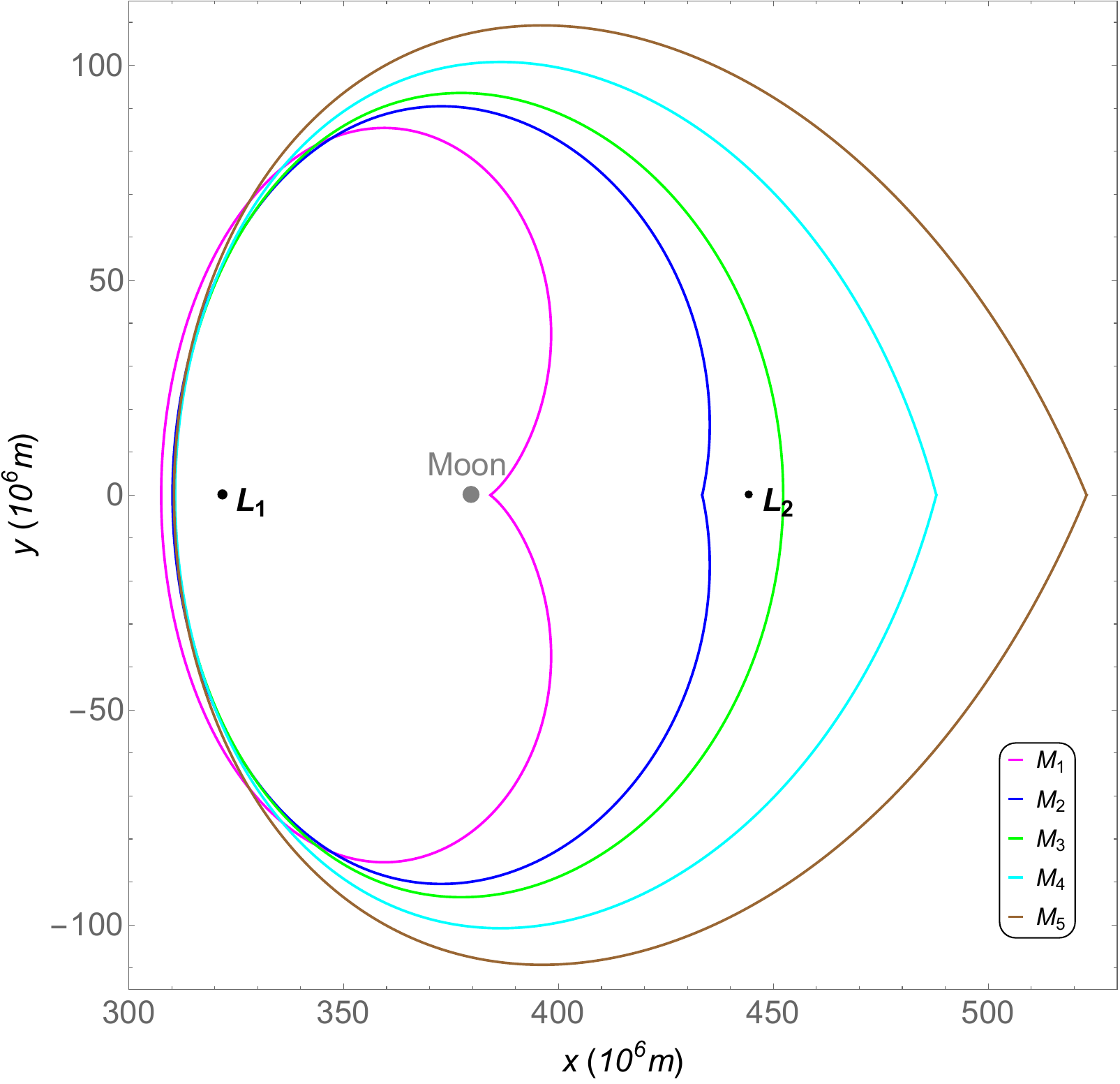}
		\caption{Several solutions for $T=13.75$ days obtained using the constrained functional shown in Eq.~(\ref{eq:cf3}). 
        The solution in green corresponds to a Distant Retrograde Orbit, which is a particular solution where $\dot{x} (0)=-\dot{x} (T)=0$.}
		\label{fig:igrl2}
\end{figure}

We demonstrate applications restricted to the planar case, where $z = 0$. We also choose the time of flight $T=13.75$ days to visualize symmetries, although other values could be selected similarly. As an initial guess, the elements of the coefficient matrix $L$ (of the free function vector expansion, see Appendix) were chosen such that the trajectory $\B{r}$ describes a circular path of radius $r_{L1}$ centered at the Lagrangian point $L_1$. Table~\ref{tab:arL1} reports solutions corresponding to a time of flight of $T = 13.75$ days for several values of $r_{L1}$. The associated trajectories are depicted in Figs.~\ref{fig:igrl1} and \ref{fig:igrl2}.

The optimization procedure converged to solutions satisfying the additional condition $\dot{x}(T/2) = 0$. As a result of this convergence, in addition to fulfilling the constraints defined in Eq.~(\ref{eq:constr2}), the following symmetries are observed: $x(0) = x(T)$, $\dot{x}(0) = -\dot{x}(T)$, and $\dot{y}(0) = \dot{y}(T)$. These properties, which are not explicitly imposed as constraints, arise naturally from the structure of the problem and the convergence behavior of the numerical method. This phenomenon aligns with known properties of the system discussed in \cite{mirror1955}.

Among the set of solutions satisfying Eq.~\eqref{eq:constr2} and converging to the specific case $\dot{x}(T/2) = 0$, certain trajectories, highlighted in red and green in Figs.~\ref{fig:igrl1} and \ref{fig:igrl2}, respectively, correspond to the special case where $\dot{x}(0) = -\dot{x}(T) = 0$. According to the mirror theorem \cite{mirror1955}, any trajectory intersecting the surface of section $y = 0$ at two points satisfying $\dot{x} = 0$ is guaranteed to be periodic. This property is illustrated in Fig.~\ref{fig:igrl1}, which highlights the connection between such periodic trajectories and the fixed point $L_1$.

Orbits with a fixed period emerge from the vicinity of $L_1$, forming a continuous family that connects the Lagrangian point to the corresponding Lyapunov orbit with the same period. This relationship is further clarified in the diagram on the right of Fig.~\ref{fig:igrl1}, which plots the position and velocity components at the initial time on the surface of section $y = 0$. These solutions are obtained through the constrained functional defined in Eq.~(\ref{eq:cf3}). Despite having a common period, the associated trajectories differ in their Jacobi constants, reflecting the underlying variation in their energy levels.

\begin{sidewaystable}
\centering
\vspace{8cm} 
    \begin{tabular}{c|l|l|c|c}    
         & \multicolumn{2}{c|}{Position ($x,y$) (m)} & \multicolumn{2}{c}{Velocity $(\dot{x},\dot{y})$ (m/s)}\rule[-7pt]{0pt}{17pt}\\ \cline{2-5}
        Name   & Initial position $\B{r}_i$ & Final position $\B{r}_f$ & Initial velocity $\B{v}_i$ & Final velocity $\B{v}_f$ \rule[-5pt]{0pt}{17pt}\\ \hline 
        $L_1$A &$(3.185283912074\times 10^8, -2\times 10^{-10})$ & $(3.185283911569\times 10^8, 4\times 10^{-10})$ & $(  35.194951526112,   -5.572243894647)$ & $(  -35.194951915596,-5.572243645276812)$ \rule[-0pt]{0pt}{12pt}\\
        $L_1$B &$(3.140548478933\times 10^8, -9\times 10^{-10})$   & $(3.140548478431\times 10^8, 0.)$             & $(  89.588656167462,  -21.876367586292)$ & $(  -89.588656538770,  -21.876367339646)$ \\
        $L_1$C &$(3.125477265453\times 10^8, -7\times 10^{-9})$   & $(3.125477264919\times 10^8, 7\times 10^{-9})$ & $( 144.222616932832,  -79.889866115870)$ & $( -144.222617321046,  -79.889865851555)$ \\
        $L_1$D &$(3.176243767321\times 10^8, 0.)$   & $(3.176243758322\times 10^8, 7\times 10^{-9})$               & $( 140.872395395421, -145.802288422431)$ & $( -140.872402169359, -145.802283786863)$ \\
        $L_1$E &$(3.217142032449\times 10^8, -1\times 10^{-8})$   & $(3.217142031251\times 10^8, 7\times 10^{-9})$ & $( 127.152719476126, -185.043999863697)$ & $( -127.152720408372, -185.043999223106)$ \\
        $L_1$F &$(3.282918899942\times 10^8, 7\times 10^{-9})$   & $(3.282918898725\times 10^8, -7\times 10^{-9})$ & $(  98.439807060059, -243.405303700152)$ & $(  -98.439808073641, -243.405302996190)$ \\
        $L_1$G &$(3.364996648482\times 10^8, 3\times 10^{-9})$   & $(3.364996651136\times 10^8, 7\times 10^{-9})$  & $(  54.488939505116, -316.307711682202)$ & $(  -54.488937035808, -316.307713437065)$ \\
        $L_1$H &$(3.448342176111\times 10^8, -1\times 10^{-8})$  & $(3.448342175847\times 10^8, -1\times 10^{-8})$ & $(  -0.000123308600, -397.425746608456)$ & $(    0.000123020876, -397.425746395866)$ \\
        $L_1$I &$(3.494056533136\times 10^8, 0.)$   & $(3.494056533433\times 10^8, 1\times 10^{-8})$               & $( -35.647361612346, -448.160321475219)$ & $(   35.647361976284, -448.160321752483)$ \\
        $L_1$J &$(3.531120980613\times 10^8, -7\times 10^{-9})$   & $(3.531120978129\times 10^8, 1\times 10^{-8})$ & $( -68.843298361216, -494.613224734647)$ & $(   68.843294943447, -494.613222048446)$ \\
        $M_1$ &$(3.841456567739\times 10^8, 3\times 10^{-8})$   & $(3.841456106279\times 10^8, -2\times 10^{-8})$  & $(1162.019989979903, -903.762348410162)$ & $(-1162.027127375701, -903.765997771761)$ \\
        $M_2$ &$(4.334399980405\times 10^8, -1\times 10^{-8})$   & $(4.334399980127\times 10^8, 1\times 10^{-8})$  & $( 105.339612046734, -478.601121345664)$ & $( -105.339611552763, -478.601121494757)$ \\
        $M_3$ &$(4.522825733181\times 10^8, -2\times 10^{-8})$   & $(4.522825734896\times 10^8, -2\times 10^{-8})$ & $(  -0.061310410539, -506.760866648326)$ & $(    0.061309899163, -506.760866774067)$ \\
        $M_4$ &$(4.879135831363\times 10^8, 1\times 10^{-8})$   & $(4.879135831329\times 10^8, 1\times 10^{-8})$   & $(-155.269026188455, -585.581665668988)$ & $(  155.269026189768, -585.581665660553)$ \\
        $M_5$ &$(5.228810302399\times 10^8, 5\times 10^{-8})$   & $(5.228810299950\times 10^8, -4\times 10^{-8})$  & $(-280.442259693692, -676.926037295773)$ & $(  280.442259687675, -676.926036561104)$ \\
    \end{tabular}
    \caption{Rectangular coordinates of the initial position $\B{r}_i$, final position $\B{r}_f$, initial velocity $\B{v}_i$, and final velocity $\B{v}_f$ of several solutions evaluated considering a common fixed time of flight of $T=13.75$ days. They were generated with different initial conditions given as a circular orbit around $L_1$ of radius $R_{L1}$.}
    \label{tab:arL1}
\end{sidewaystable}

\subsubsection{Narrowing solutions by adding constraints}

Although the constraints introduced in Sec.~\ref{sec:exact_surface} are sufficient to describe a first recurrence, which is valuable for investigating general system properties, the resulting solution set may not be unique and can vary depending on the underlying dynamical model (i.e., the equations of motion). In this regard, the trajectories shown in Figs.~\ref{fig:igrl1} and \ref{fig:igrl2} are not solely determined by the constraints defined in Eq.~(\ref{eq:constr2}), but also emerge as outcomes of the optimization procedure.

To isolate particular types of solutions, additional constraints can be imposed to restrict the solution space. This approach allows for the targeting of specific “solutions of interest” by appropriately formulating constraints and deriving a corresponding constrained functional. For example, imposing the constraint $\dot{x}(T/2) = 0$ and selecting the support function component $s_x = t$ yields the following constrained functional:
\begin{equation}\label{eq:cf4}
    x(t) = g_x(t) - \dot{g}_{T/2} t,
\end{equation}
where $g_x(t)$ denotes the $x$-component of the free function, and $\dot{g}_{T/2}$ represents its derivative evaluated at $t = T/2$. In this formulation, the optimization procedure is guided to converge to trajectories satisfying $\dot{x}(T/2) = 0$, such as those depicted in Figs.~\ref{fig:igrl1} and \ref{fig:igrl2}.

The solution space can be further narrowed by introducing additional constraints. For instance, specifying the initial $x$-position via the constraint $x(0) = x_0$ enables selection of a particular $x$-coordinate on the surface of section. Setting $x_0 = 4.334399980405431 \times 10^8$ leads to convergence toward the solution labeled $M_2$ in Table~\ref{tab:arL1} and shown in Fig.~\ref{fig:igrl2}. Alternatively, enforcing $x(0) = 3.185283912074594 \times 10^8$ results in at least two distinct solutions, as illustrated in the right panel of Fig.~\ref{fig:igrl1}.

Thanks to the mirror theorem \cite{mirror1955}, it is not necessary to explicitly impose the terminal constraint $x(T) = x_0$ when both $\dot{x}(T/2) = 0$ and $x(0) = x_0$ are enforced. The symmetry properties of the system ensure that this condition is inherently satisfied.

\section{Discrete time analysis}
\label{sec:strob}

A \textit{stroboscopic map} acting on the phase space can be defined with the transformation
$$
    (\B{r}(t), \dot{\B{r}}(t)) \rightarrow (\B{r}(t+T), \dot{\B{r}}(t+T)),
$$
where, without loss of generality, the initial time is set to $t = 0$ to align with the formulation adopted in this section. This definition enables a reduction in the order of the system by one, effectively transforming a continuous-time system into an autonomous discrete-time system.

In addition to identifying crossings of a surface of section, such as those defined in Eqs.~(\ref{eq:surface_def}) or (\ref{eq:surfacey}), the procedure presented here can also be used to discretize time. This is particularly useful in cases where the original system is non-autonomous and subjected to a periodic forcing term with period $T/\lambda$, where $\lambda \in \mathbb{N}^+$ is the period of the periodic orbit represented in the \textit{stroboscopic map}. Since the time interval $T$ is specified, the resulting discrete map incorporates both spatial and temporal crossings, allowing for analyzing the system’s evolution in a unified treatment with two dimensions reduced (considering $t$ as a dimension variable).

\subsection{Application to the Bicircular Biplanar 4BP}
\label{sec:appl4BP}

In comparison to the Circular Restricted Three-Body Problem (CRTBP) defined in Eq.~(\ref{eq:crtbp}), the perturbative effect of the Sun, denoted by $\B{p}_s$, is introduced through the following expression \cite{allansr}:
\begin{eqnarray}\label{eq:ps}
    \B{p}_s = -\frac{\mu_s}{r_s^3}\B{r}_s - \frac{\mu_s}{R_s^3}\B{R}_s,
\end{eqnarray}
where $\mu_s$ is the gravitational parameter of the Sun, and $\B{r}_s = \B{r} - \B{R}_s$. The position vector of the Sun in the rotating frame, $\B{R}_s$, is defined as,
\begin{equation*}
    \B{R}_s = R_s \, \begin{Bmatrix} \cos (\omega_s t + \gamma), &  \sin (\omega_s t + \gamma), & 0\end{Bmatrix}\T
\end{equation*}
where $\omega_s$ represents the angular velocity of the Sun relative to the Earth-Moon rotating frame, and $\gamma$ is a phase constant specifying the initial position of the Sun within this frame. The bicircular, biplanar four-body problem (4BP) is thus defined by augmenting the right-hand side of Eq.~(\ref{eq:crtbp}) with the perturbative acceleration given in Eq.~(\ref{eq:ps}).

It is important to emphasize that the inclusion of this perturbation renders the system non-autonomous. Specifically, the force term in Eq.~(\ref{eq:ps}) is periodic with period $2\pi / \omega_s$. To facilitate the analysis of periodic orbits, i.e., temporally repeating solutions, within a time-discretized framework, we set the total integration time to $T = 2\pi / \omega_s$, and impose the following boundary conditions in Cartesian coordinates:
\begin{equation}\label{eq:constr3}
    \begin{cases} y(0) = 0, \\ y(T) = 0, \\ \dot{x}(0) = 0, \\ \dot{x}(T) = 0.\end{cases}
\end{equation}
In addition to the constrained functional for the $y$-component presented in Eq.~(\ref{eq:cf3}), the constrained functional for the $x$-component can be derived by using the support functions $t$ and $t^2$, following the procedure outlined in earlier sections. The resulting expression is given by:
\begin{equation}\label{eq:cf3y}
    x(t) = g_x(t) - \dot{g}_x(0) ~ t + \big( \dot{g}_x(0) - \dot{g}_x(T) \big)\dfrac{t^2}{2T},
\end{equation}
where $\dot{g}_x(0)$ and $\dot{g}_x(T)$ denote the derivatives of the $x$-component of the free function $\B{g}$ evaluated at $t = 0$ and $t = T$, respectively.

\begin{figure}
\centering\includegraphics[scale=0.38]{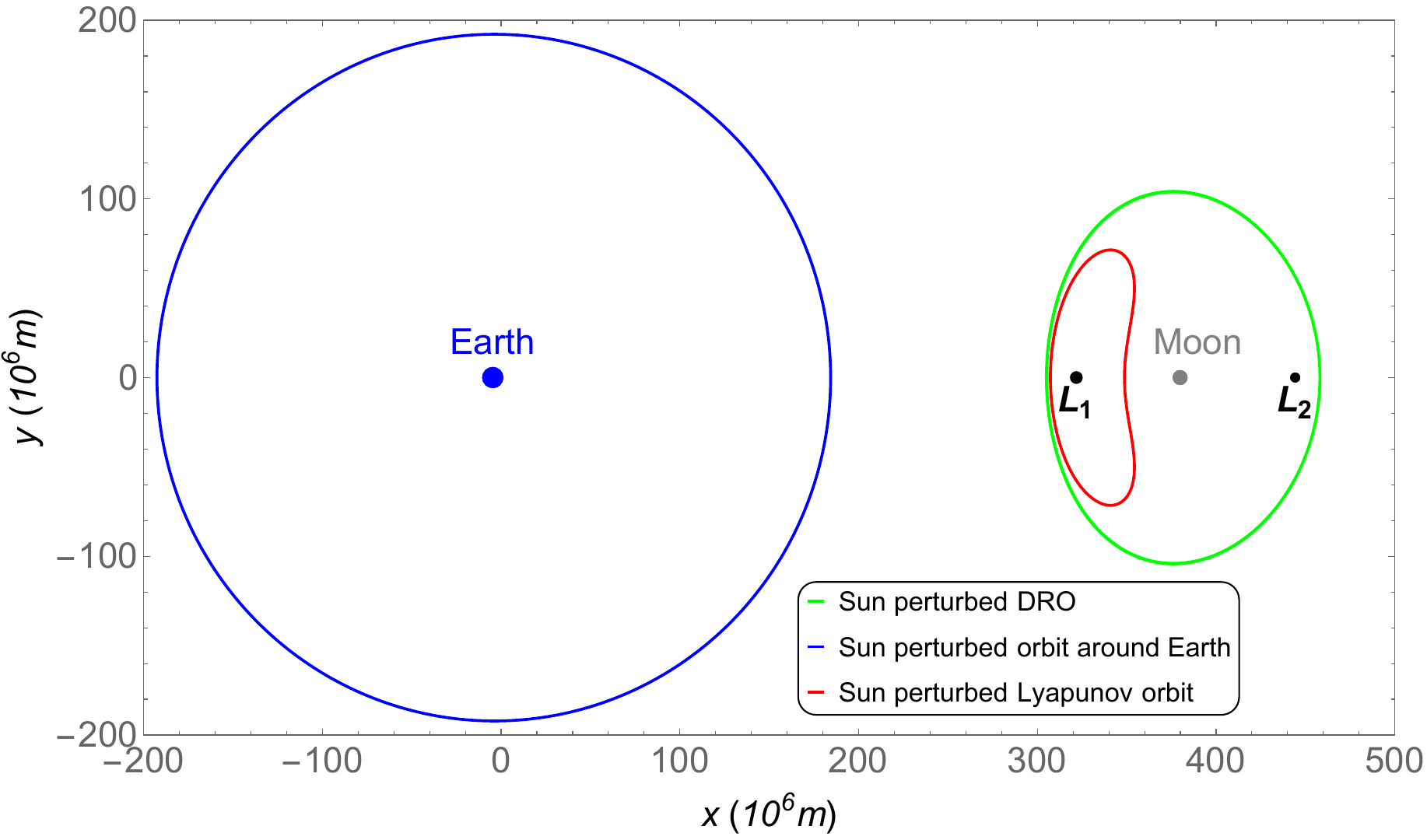}
\caption{Planar perturbed periodic solutions in the biplanar bicircular 4BP: Lyapunov orbits are shown in red; Distant Retrograde Orbits (DROs) around the Moon are shown in green; and Earth-centered periodic orbits are depicted in blue. The period of these orbits is 29.52887871613042 days, which corresponds to the period of the Sun's motion (and its gravitational influence on the spacecraft) in the Earth-Moon rotating frame. These solutions were obtained via numerical convergence using the constrained functionals shown in Eqs.~\eqref{eq:cf3} and \eqref{eq:cf3y}.}
\label{fig:lyap4bp}
\end{figure}

We set the phase of the Sun to $\gamma = 0$ and focus on applications to the planar case, where $z = 0$, as was done in the previous sections to simplify the analysis. Planar perturbed periodic orbits with period $T = 2\pi/\omega_s$ (corresponding to 29.52887871613042 days) are shown in Fig.~\ref{fig:lyap4bp}. These trajectories are periodic under the influence of the Sun's gravitational perturbation. The Sun is fixed in a \textit{stroboscopic map}.

It is important to emphasize that the perturbed Distant Retrograde Orbits (DROs) and Lyapunov orbits become periodic with respect to the surface of section defined in Eq.~\eqref{eq:surfacey} only after the second crossing; that is, their effective period with respect to the section is 2. Consequently, their shape and distance from their center are comparable to those of the corresponding unperturbed orbits over a half-period, $T/2$.

The coordinates of these periodic orbits are provided in Table~\ref{tab:IC4BP}. 
These solutions are computed using the constrained functionals defined in Eqs.~\eqref{eq:cf3} and \eqref{eq:cf3y}, ensuring that the boundary conditions are exactly satisfied. 
The associated errors shown in this table are defined as follows. We integrate the initial positions and velocities shown in Table~\ref{tab:IC4BP} for the respective period of the periodic orbit using the initial value problem (IVP) approach described in Sec.~\ref{sec:IVP}, in which the boundary conditions are not enforced. The absolute value of the difference between the final and initial positions and velocities components are the errors summarized in  Table~\ref{tab:IC4BP}.

The maximum deviations between the constrained and IVP-based trajectories are on the order of $10^{-2}$ m in position and $10^{-7}$ m/s in velocity, based on the errors shown in Table~\ref{tab:IC4BP}. These results confirm that the computed orbits are highly accurate and effectively periodic.

In general, the gravitational perturbation from the Sun is significant, and neglecting it can lead to substantial deviations in trajectory predictions. However, analyzing such periodic solutions under solar perturbation is inherently challenging, as known periodic orbits and their associated invariant structures in the CRTBP may change drastically over time. The methodology introduced in this section provides a practical framework for addressing this challenge, enabling the analysis of periodicity through both spatial crossings with a surface of section and temporal discretization. This approach supports the construction not only of Poincaré maps but also of \textit{stroboscopic maps}, acting in phase space as needed.

\begin{table}[ht]
    \centering
    \resizebox{1.2\textwidth}{!}{
    \begin{tabular}{c|c|c}    
        Orbit   & Position ($x,y$) (m)  & Error ($x,y$) (m)   \rule[-5pt]{0pt}{17pt}\\ \hline
        DRO &$(305043082.71037555, -2.9802322387695312\times 10^{-8})$ &  $( 1.93715\times 10^{-5},  3.36708\times 10^{-5})$\rule[-0pt]{0pt}{12pt}\\
        AE &$(-192525720.62106377, 4.470348358154297\times 10^{-8})$ &   $( 5.66244\times 10^{-7},  8.61005\times 10^{-6})$   \\
        Lyapunov &$(348714288.19374937, 2.60770320892334\times 10^{-8})$ & $( 1.10705\times 10^{-2},  2.01820\times 10^{-2})$ \\ \hline\hline
        Orbit   & Velocity $(\dot{x},\dot{y})$ (m/s)  & Error $(\dot{x},\dot{y})$ (m/s)  \rule[-5pt]{0pt}{17pt}\\ \hline
        DRO & $(-1.1368683772161603\times 10^{-12}, 549.9860059765858 )$   & $( 1.78716\times 10^{-11},  7.24185\times 10^{-11})$ \rule[-0pt]{0pt}{12pt}\\
        AE &$(-2.6858515411731787\times 10^{-12}, -965.3812911936908)$     & $(6.42270 \times 10^{-11},  5.22959\times 10^{-12})$     \\
        Lyapunov &$(6.821210263296962\times 10^{-12}, -458.3080802645622)$  & $( 2.13067\times 10^{-7}, 1.05525 \times 10^{-7})$  \\ \hline
    \end{tabular}}
    \caption{Rectangular coordinates of the periodic perturbed orbits with period $2 \pi /  \omega_s$ shown in Fig.~\ref{fig:lyap4bp} and their associated errors.}
    \label{tab:IC4BP}
\end{table}

\section*{Conclusions}

In this work, we presented techniques for constructing first recurrence maps, also known as Poincar\'e maps. The first approach involves formulating the problem as an initial value problem (IVP), where the map is traditionally generated by propagating a selected grid of initial conditions. The proposed method leverages the efficiency of the TFC to evaluate the intersection of each trajectory, originating from the initial grid, with the surface of section.

This method naturally yields a continuous constrained functional, which is then employed to solve for the root corresponding to the intersection with the surface of section. For simple forms of the free function, the root can be computed analytically; otherwise, a Newton method is employed for numerical evaluation. This approach is significantly more efficient than traditional numerical integration with progressively smaller time steps to detect surface crossings.
Importantly, the technique eliminates the need for external interpolation procedures, as the constrained functional inherently provides a continuous solution, constructed through a highly efficient blend of analytical and numerical components, capable of achieving machine-level precision with modest computational effort. Additionally, it obviates the need for variable transformation at each crossing.

The problem is also formulated as a boundary value problem (BVP), allowing for the identification of periodic solutions, as demonstrated in the context of the Circular Restricted Three-Body Problem (CRTBP). Furthermore, the notion of partially periodic solutions is introduced for cases in which only a subset of variables exhibits periodicity. This type of symmetry enables the exact computation of first recurrences.

The technique is then developed for time-dependent systems. The technique allows for further decrease in the dimension of the system by analyzing it in a time-discrete domain. Such a methodology is very useful to analyze systems subject to periodic forces.

Overall, the approach allows for the refinement of surface-of-section crossings by enforcing additional constraints, enabling the targeted search for specific solutions of interest with high accuracy and computational efficiency.

\section*{Acknowledgments}

AKAJ thanks T. Vaillant for elucidating discussions on the subject of this study.

\section*{Fundings}

This research was done with own resources.

\section*{Contributions}

AKAJ: conceptualization, writing - first version, methodology, investigation, formal analysis, and software. DM: conceptualization, formal analysis, and writing - review \& editing.


%
%

\bibliographystyle{unsrt}
\bibliography{_references}

\section*{Appendix}

\subsection*{Numerical procedure}
\label{sec:numerical}

The constrained functional given in Eq.~(\ref{eq:cf2}), (\ref{eq:ce2c}), or (\ref{eq:cf3}), along with its derivative, is substituted into the differential equation in Eq.~(\ref{eq:eqmotion}). This substitution results in the following unconstrained differential equation:
\begin{equation}\label{eq:unconst}
    \ddot{\B{g}}(t) - \B{a}'(\B{g}, \dot{\B{g}}, \ddot{\B{g}}, t) = \B{0},
\end{equation}
where, for example, in the case of the constrained functional defined by Eq.~(\ref{eq:cf2}), the modified specific force $\B{a}'$ is given by
\[
\B{a}' = \B{a}(\B{r}(\ddot{\B{g}}, \dot{\B{g}}, \B{g}, t), \dot{\B{r}}(\ddot{\B{g}}, \dot{\B{g}}, \B{g}, t), t).
\]

The free function $\B{g}(t)$ is expressed as
\begin{equation}\label{eq:freefunction}
    \B{g}(t) = L \, \B{h} (\tau(t)),
\end{equation}
where $L$ is a constant matrix of unknown coefficients of dimension $3 \times (m - k + 1)$, and $\B{h}(\tau)$ is a $(m - k + 1) \times 1$ vector composed of orthogonal basis functions, such as Chebyshev polynomials of the first kind \cite{abramowitzstegun1972}. Here, $m$ denotes the highest degree (i.e., truncation order) of the polynomial basis, and $k$ is the number of constraints applied. The basis elements in $\B{h}$ are selected to be linearly independent of the support functions, resulting in a total of $(m - k + 1)$ terms.

Since Chebyshev polynomials are defined on the interval $-1 \leq \tau \leq 1$, a change of the independent variable is required:
\[
\tau = \frac{2t}{T_s} - 1,
\]
to map the time interval $[0, T_s]$ into the standard domain of the polynomials.

The time domain is discretized into $N + 1$ points using Chebyshev-Gauss-Lobatto nodes, as described in \cite{lanczos1988applied}:
\begin{equation}\label{eq:distr}
    t_j - t_0 = \left(1 - \cos\left(\frac{j \pi}{N}\right)\right) \frac{T_s}{2}, \quad \text{for } j \in \llbracket 0 : N \rrbracket.
\end{equation}

By combining the free function representation from Eq.~(\ref{eq:freefunction}) with the time discretization in Eq.~(\ref{eq:distr}), the differential equation in Eq.~(\ref{eq:unconst}) can be transformed into the discrete system of equations:
\begin{equation}\label{eq:eqmotion3}
    \B{a}''(L) = \B{0},
\end{equation}
where $\B{a}'': \mathbb{R}^{3 \times (m - k + 1)} \to \mathbb{R}^{3 \times (N + 1)}$ represents the residual function evaluated at the collocation nodes.

The solution to the resulting system of $3 \times (N + 1)$ nonlinear equations, with $3 \times (m - k + 1)$ unknowns corresponding to the elements of $L$, is obtained using a nonlinear least squares optimization method \cite{NDE}. This method minimizes the norm of the residual matrix $\B{a}''$.

Once the optimal matrix $L$ is found, the free function $\B{g}(t)$ can be reconstructed using Eq.~(\ref{eq:freefunction}). Substituting this free function back into the constrained functional in Eq.~(\ref{eq:cf2}) yields the complete solution $\B{r}(t)$ of the original system of differential equations in Eq.~(\ref{eq:eqmotion}), incorporating the specified form of the specific force $\B{a}(\B{r}, \dot{\B{r}}, t)$.


\end{document}